# Automatic reorientation by deep learning to generate short-axis SPECT myocardial perfusion images

Fubao Zhu[1], Guojie Wang[1], Chen Zhao[2], Saurabh Malhotra[3,4], Min Zhao[5], Zhuo He[2], Jianzhou Shi[6], Zhixin Jiang[6*], Weihua Zhou[2,7*]

1. School of Computer and Communication Engineering, Zhengzhou University of Light Industry, Zhengzhou, Henan, 450000, China

2. Department of Applied Computing, Michigan Technological University, Houghton, MI, 49931, USA

3. Division of Cardiology, Cook County Health and Hospitals System, Chicago, IL, 60612, USA

4. Division of Cardiology, Rush Medical College, Chicago, IL, 60612, USA

5. Department of Nuclear Medicine, Xiangya Hospital, Central South University, Changsha, 410008, China

6. Department of Cardiology, The First Affiliated Hospital of Nanjing Medical University, Nanjing, 210000, China

7. Center for Biocomputing and Digital Health, Institute of Computing and Cybersystems, and Health Research Institute, Michigan Technological University, Houghton, MI, 49931, USA

*Address for correspondence*

Weihua Zhou, PhD

E-mail: whzhou@mtu.edu

Tel: 1-906-487-2666

Address: 1400 Townsend Drive, Houghton, MI 49931, USA

Or

Zhixin Jiang, MD, PhD

E-mail: zhixin_jiang@njmu.edu.cn

Tel. +86-13913832207

Address: Guangzhou Road 300, Nanjing, Jiangsu, China 210029

***Acknowledgements***

This research was supported in part by a Michigan Technological Research Excellence Fund Research Seed grant (PI: Weihua Zhou), a seed grant from Michigan Technological University Health Research Institute (PI: Weihua Zhou), Henan Science and Technology Development Plan 2022 (Project Number: 222102210219), and the National Natural Science Foundation of China under Grant 62106233.

**ABSTRACT**

Single photon emission computed tomography (SPECT) myocardial perfusion images (MPI) can be displayed both in traditional short-axis (SA) cardiac planes and polar maps for interpretation and quantification. It is essential to reorient the reconstructed transaxial SPECT MPI into standard SA slices. This study is aimed to develop a deep-learning-based approach for automatic reorientation of MPI. **Methods:** A total of 254 patients were enrolled, including 228 stress SPECT MPIs and 248 rest SPECT MPIs. Five-fold cross-validation with 180 stress and 201 rest MPIs was used for training and internal validation; the remaining images were used for testing. The rigid transformation parameters (translation and rotation) from manual reorientation were annotated by an experienced operator and used as the ground truth. A convolutional neural network (CNN) was designed to predict the transformation parameters. Then, the derived transform was applied to the grid generator and sampler in spatial transformer network (STN) to generate the reoriented image. A loss function containing mean absolute errors for translation and mean square errors for rotation was employed. A three-stage optimization strategy was adopted for model optimization: 1) optimize the translation parameters while fixing the rotation parameters; 2) optimize rotation parameters while fixing the translation parameters; 3) optimize both translation and rotation parameters together. **Results:** In the test set, the correlation coefficients of the translation distances and rotation angles between the model prediction and the ground truth were 0.99 in X axis, 0.99 in Y axis, 0.99 in Z axis, 0.99 along X axis, 0.99 along Y axis and 0.99 along Z axis, respectively. For the 46 stress MPIs in the test set, the Pearson correlation coefficients were 0.95 in scar burden and 0.95 in summed stress score; for the 46 rest MPIs in the test set, the Pearson correlation coefficients were 0.95 in scar burden and 0.95 in summed rest score. **Conclusions:** Our deep learning-based LV reorientation method is able to accurately generate the SA images. Technical and clinical validations show that it has great promise for clinical use.

**Key Words:** SPECT MPI; reorientation; deep learning; convolutional neural networks

**Abbreviations**

| | |
|---|---|
| SPECT | single photon emission computerized tomography |
| MPI | myocardial perfusion imaging |
| CNN | convolutional neural networks |
| STN | spatial transformer network |
| LV | left ventricular |
| SA | short-axis |
| ECTb | Emory Cardiac Toolbox |

## INTRODUCTION

Myocardial perfusion imaging (MPI) with single photon emission computed tomography (SPECT) is regarded as one of the most utilized non-invasive cardiac imaging modalities in the diagnosis of coronary artery disease(*1*). The display of SPECT MPI in the transaxial view is patient-specific because of the differences in left-ventricular (LV) orientation between patients, which complicates the visual interpretation of images(*2*). In clinical use, LV transaxial images after 3D reconstruction usually need to be reoriented to the standard short-axis (SA) view(*3,4*). Moreover, the accuracy of identification of regional perfusion defects in the LV myocardium can be improved by reorienting the LV into SA(*5*). In SPECT MPI, it has become routine to reorient cardiac images acquired in the transaxial axis to the standard SA view(*6-9*).

Traditional manual reorientation is time-consuming and less reproducible(*10*). A number of studies have shown that significant artifacts, and quantitative analysis may be misleading due to incorrect reorientation(*11,12*). Therefore, precise and automated methods are needed to obtain satisfactory SA images and reliable quantitative results(*13*). Automated methods for reorientation have been investigated and used in clinical routine(*14-16*). Nevertheless, they rely on the integrity of the myocardium(*17*). Therefore, current commercial software also provides manual alternatives to prevent automated methods from failing in reorientation as well. Recently, deep learning has shown great performance in medical image processing, including tissue and organ localization and reorientations(*17-20*). Zhang et al(*17*). developed a method using convolutional neural networks (CNN) for automatic reorientation of MPI. It achieved a high performance, but further improvement is still needed for clinical use.

This study is aimed to develop a deep-learning-based approach for automatic reorientation of MPI and evaluate its values in MPI quantitative analysis for clinical use.

**MATERIALS AND METHODS**

**Data Acquisition**

We retrospectively enrolled 254 patients (226 stress and 247 rest MPIs) from the First Affiliated Hospital of Nanjing Medical University. Both rest and stress ECG-gated SPECT MPIs were performed approximately 60 min after injection of 20-30 mCi Tc-99m sestamibi. The MPI images were acquired on a dual-headed camera (CardioMD, Philips Medical Systems) with a standard protocol. The imaging parameters included a 20% energy window around 140 KeV, 180°orbit, 32 steps with 25 seconds per step, 8-bin gating, and 64 planar projections per gate. All SPECT planar images were reconstructed by Emory Reconstruction Toolbox (ERTb v2.0; Syntermed, Atlanta, GA) with 3 iterations and 10 subsets of ordered subset expectation maximization, and then low-pass filtered with Butterworth at a cutoff frequency of 0.3 cycles/mm to order 10 subsets.

The image size was 64 * 64 in each slice, LV slice numbers range from 22 to-36, and the voxel size was 6.4mm × 6.4mm × 6.4mm. All ungated reconstructed transaxial images were reoriented into SA images by experienced operators using Emory Reconstruction Toolbox (ERTb v2.0; Syntermed, Atlanta, GA). The six parameters for translation and rotation, as shown in [Figure 1](), were identified using an 3D image processing software Elastix (*21,22*) to represent the transformation between the transaxial images and SA images. The deidentified dataset is available at https://github.com/MIILab-MTU/SPECTMPILVReorientation.

Ninety percent of the stress and rest MPIs were used for the five-fold cross-validation, 180 stress and 201 rest SPECT MPIs were used for training and internal validation, and the remaining images were used for testing and clinical evaluation. Institutional review board approval by the Institutional Ethical Committee of the First Affiliated Hospital of Nanjing Medical University was obtained with no informed consent required for this HIPAA-compliant retrospective analysis.

**Data augmentation**

To increase the sample size and avoid overfitting, several image processing techniques were applied. First, a Gaussian distribution was calculated for all the transformation parameters. Secondly, each set of data generates an additional 40 sets of transformation parameters from the truncated Gaussian distributions. Finally, SA images were inversely transformed with Gaussian-generated parameters to generate new images, regarded as the augmented transaxial images.

After data augmentation, a total of 15621 MPIs were available for training and cross-validation.

**Deep-learning-based method**

Our 3D network architecture for LV reorientation is shown in [Figure 2](). All 3D MPIs were rescaled to 64*64*32 to facilitate the network training and prediction. To enhance the contours in the SPECT MPI, gradient images of the transaxial images were calculated, and combined with the transaxial images to form a two-channel network. As a result, the network consists of two channels (2 * 64 * 64 * 32) as the input for the learning process.

Three reorientation blocks using CNNs, as shown in [Figure 2]() Top, were employed to predict transformation parameters for image reorientation. The CNN architecture is illustrated in [Figure 2]() Bottom. A spatial transformer network (STN)(*23*) was used after each reorientation block to generate the reoriented images. STN has been used in our previous studies to combine results predicted by CNN and prior knowledge for improved image segmentation (*24,25*). It provides three sub-modules: a predicted network of transformation matrices, a grid generator for coordinate mapping, and a sampler for pixel collection. As shown in [Figure 2]() Top, skip links are transformations of images through samplers.

Mean absolute error (MAE) and mean square error (MSE) was used in the loss function of translation parameters and rotation parameters, respectively, as shown in Eq. 1 and Eq. 2.

$$L_{MSE}(\hat{y}, y) = \frac{1}{n}\sum_{i=1}^{n}(\hat{y}_i - y_i)^2, \quad \text{Eq. 1}$$

$$L_{MAE}(\hat{y}, y) = \frac{1}{n}\sum_{i=1}^{n}|\hat{y}_i - y_i|, \quad \text{Eq. 2}$$

Where $\hat{y}_i$ and $y_i$ are the predicted and ground truth and n indicates the number of parameters, i.e., n = 3 for translation and rotation parameters.

To train both translation parameters and rotation parameters, a composite loss function that combines MAE with MSE was used:

$$L_{par}(\hat{y}, y) = \mu L_{MSE}(\hat{y}, y) + L_{MAE}(\hat{y}, y), \quad \text{Eq. 3}$$

where μ is an empirical weight, which balances the losses from MSE and MAE.

Both stress and rest MPIs were input into the network. Our proposed method was implemented in Python using PyTorch, and the model was trained on a workstation with an NVIDIA V100 GPU. An Adam Optimizer was used to fine-tune the weights of the network. The model was trained for 200 epochs with a batch size of 8. We trained only the translation parameters in the first 80 epochs, only rotation parameters in the 81-160 epochs, and both translation and rotation parameters in the 161-200 epochs.

**Evaluation and Statistical Analysis**

Both technical and clinical evaluations were conducted in the test set with 46 stress and 46 rest MPIs. To technically evaluate the transformation parameters, the correlation was reported. To clinically validate the accuracy of clinical parameters measured from the short-axis images generated, both correlation and MAE were reported. All clinical parameters were measured by a commercial software package (Emory Cardiac Toolbox 4.0; Atlanta, GA).

To further confirm the effectiveness of STN in the LV reorientation, both 3DVGG and

3DResNet were implemented as the benchmark, and further we incorporated the three-stage optimization strategy with 3DVGG and 3DResNet to validate their values in optimizing the transformation parameters. For both of them, the first 80 epochs were used for training translation parameters, the 81-160 epochs for training rotation parameters, and 160-200 epochs for overall training. For 3DVGG and 3DResNet without the three-stage optimization, a total of 160 epochs were used for training translation and rotation parameters together. The same loss function as in our method was used for the training.

**RESULTS**

Table 1 shows the correlations between the predicted reorientation parameters by different models and the ground truth, the number of networks, and the running/test time. Correlation coefficients in our method were better than those by 3DVGG, 3DResNet, CNN+STN, and 3DVGG-ThreeStages. Compared to 3DResNet-ThreeStages, our method achieved almost the same accuracy but it required a much smaller number of network parameters and ran faster for training and testing.

Figure 3 shows the quantitative analysis of transformation parameters. There were excellent correlations in the reorientation parameters between our method and the ground truth. The correlation coefficients were 0.994, 0.996, 0.996 for the translation parameters and 0.992, 0.997, 0.998 for the rotation parameters (all P values < 0.01).

Figure 4 shows three examples by the visualization of three orthogonal orientations. The images obtained by our method were similar to those in the ground truth.

**Clinical quantitative evaluation**

Figure 5 shows high agreement of MPI quantitative parameters between our method and the ground truth. The correlation coefficients of both scar burden and summed score for stress/ rest MPIs were all 0.95 (all P values < 0.001). The MAEs of scar burden and summed score were 0.0044 and 0.35 for stress MPIs, and 0.0053 and 0.32 for rest MPIs, respectively.

Figure 6 shows the polar maps of three stress MPIs (top, normal; middle and bottom, mild perfusion defect) and three rest MPIs (top, normal; middle, mild perfusion defect; bottom, severe perfusion defect). The scar burden and summed stress/rest scores of 17-segment perfusion defect show strong consistency between the ground truth and our method.

**DISCUSSION**

In this study, we developed a deep learning network for LV reorientation on SPECT MPI to generate short-axis images. The correlations of all transformation parameters between our prediction and the ground truth are strong. Furthermore, our clinical evaluation for the quantification of MPI shows strong correlations for clinical parameters between the network generated images and ground truth.

The traditional methods require the extraction of LV myocardial contours (*14,15*) or localization of LV apex and base(*13*) before LV reorientation. This may lead to failure when the LV contours are not clear. However, the reorientation can be accomplished without extracting contours in our method. In addition, several reasons which may cause the failure of LV reorientation were mentioned(*14,15,17*): 1) the reorientations angles exceeding 45 degrees; 2) severe overlap between LV and other organs. In a study by Germano et al 6 of 400 groups failed(*15*) while 8 of 124 groups failed in the study by Mullick et al(*14*). These problems were alleviated or eliminated in our method, as illustrated in Fig. 7.

There are few studies using deep learning for LV reorientation by far. In the recent study by Zhang et al(*17*), the correlation coefficients were 0.928, 0.958, 0.994 for translation, and the correlation coefficients were 0.973, 0.96, 0.970 for rotation. When tested with our dataset, the correlation coefficients were 0.987, 0.99, 0.995 for translation, and the correlation coefficients were 0.981, 0.992, 0.987 for rotation. The differences of the prediction accuracy confirm two major innovations in our method: 1) we used the combination of MAE and MSE in the loss function to train the translation and rotation parameters; 2) we designed an effective three-stage optimization for these transformation parameters. As shown in Table 1, the three-stage optimization significantly improved the performance for all the three networks (3DVGG, 3DResNet, and CNN+STN).

This study shows our continuous effort in MPI image analysis. Our method has received rigorous technical and clinical validations and the results demonstrated excellent performance. Combined with our LV segmentation methods(*26-28*), we will further integrate it with our approaches for MPI functional quantification(*9,29,30*), and image fusion between cardiac functions from MPI and coronary anatomy from invasive coronary angiograms/ venograms(*31-34*).

**Limitations**

This study enrolled a relatively small number of patients from a single medical center with the inherent limitation of such study design.

**NEW KNOWLEDGE GAINED**

We have developed a deep-learning-based method for LV reorientation on SPECT MPI. It can generate accurate SA images in a short time. There are high correlations for MPI quantification parameters measured from short-axis images generated by manual reorientation and by automatic reorientation using our model.

**CONCLUSION**

Our deep learning-based method is able to accurately generate the SA images. Technical and clinical validations show that it has great promise for clinical use.

**DISCLOSURES**

All authors declare that there are no conflicts of interest.

**KEY POINTS**

QUESTION: Can deep learning be applied to generate short-axis slices from transaxial images for automated SPECT MPI image analysis?

PERTINENT FINDINGS: Our deep learning-based method accurately generated the short-axis slices in a fast running speed.

IMPLICATIONS FOR PATIENT CARE: Our deep learning-based method has great promise in clinical use for automated SPECT MPI quantification.

Figures

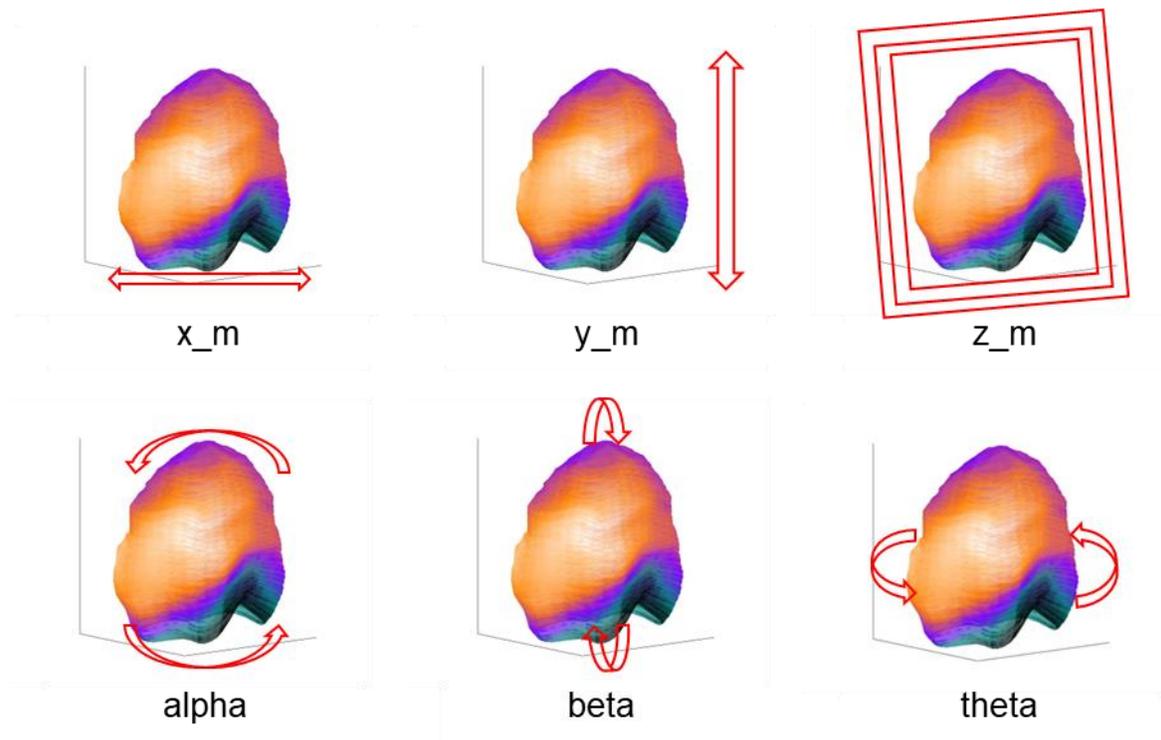

**FIGURE 1.** The six transformation parameters for left-ventricular reorientation to generate short-axis images. Top: x_m, y_m, z_m for translation. Bottom: alpha, beta, theta for rotation.

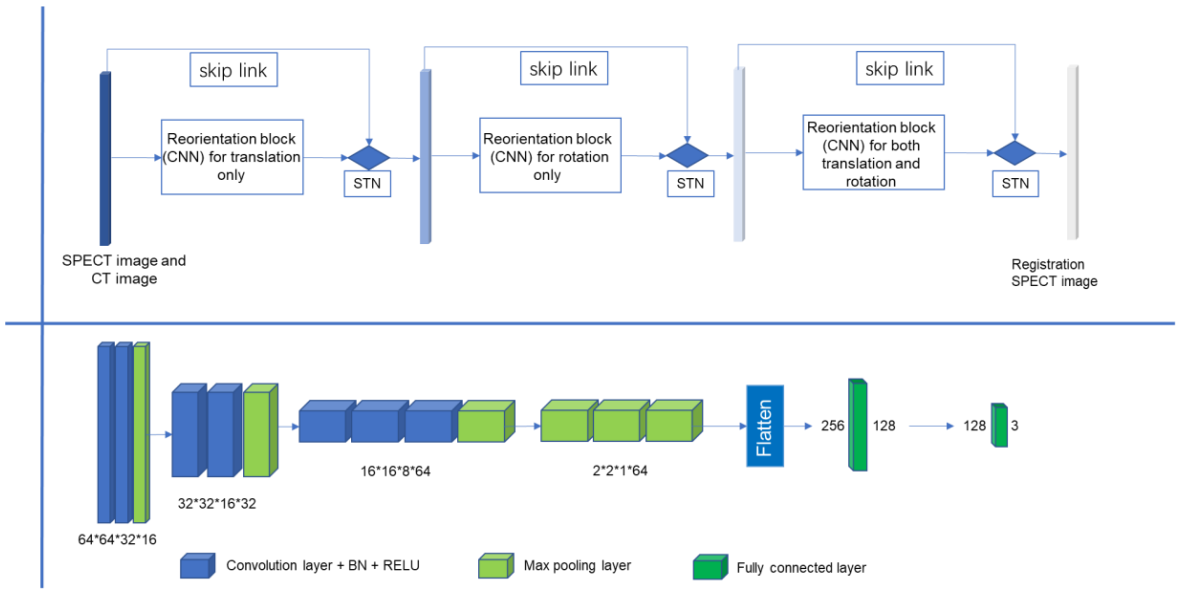

**FIGURE 2.** The optimization of transformation parameters by deep learning for left-ventricular reorientation. Top: Combination of reorientation blocks and STN for image reorientation. Bottom: 3D CNN architecture in each reorientation block to calculate the transformation parameters. CNN, convolutional neural network; STN, spatial transformer network; BN, batch normalization; RELU, rectified linear unit.

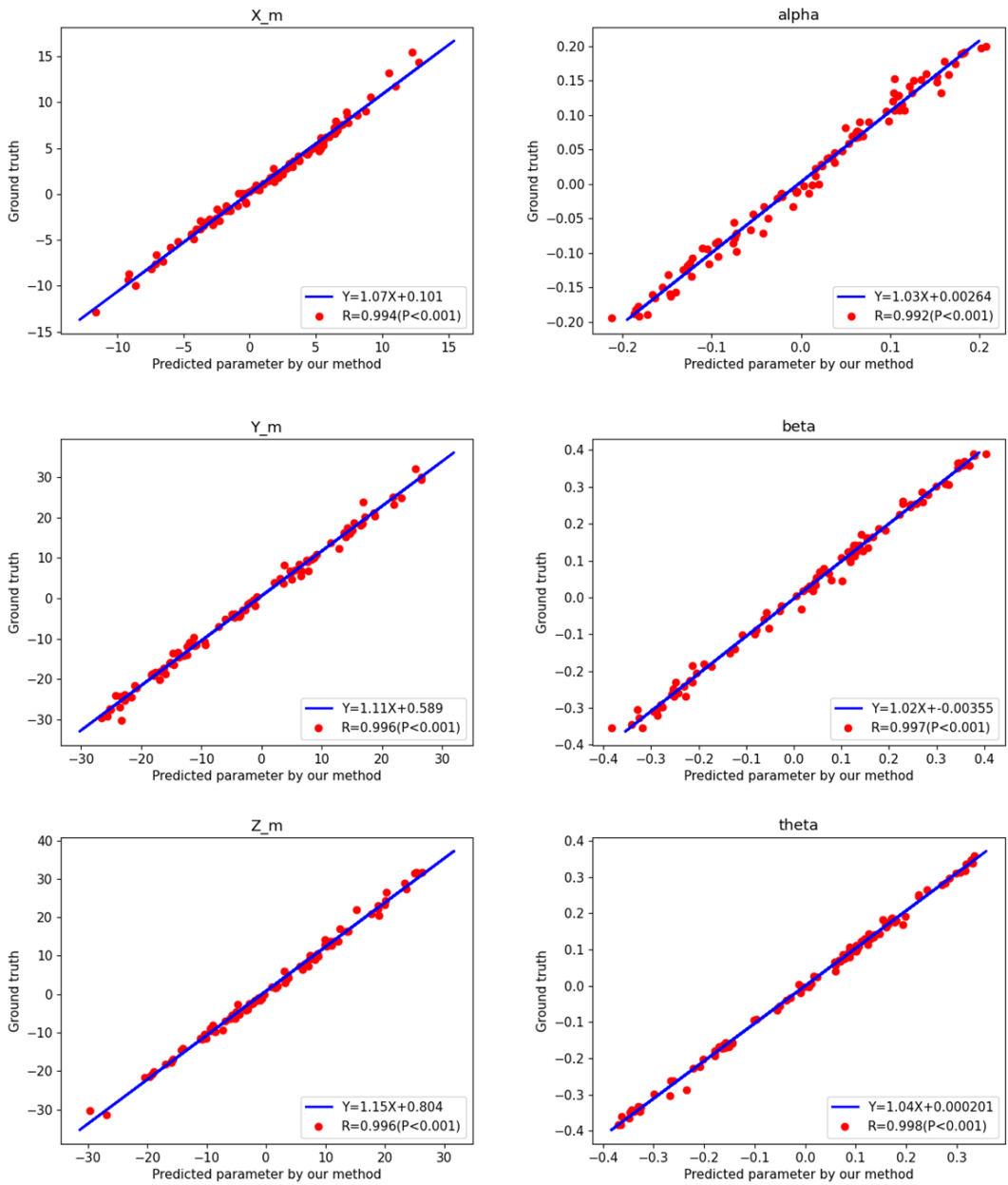

**FIGURE 3.** Linear regression analysis to evaluate the translation parameters (left column) and rotation parameters (right column) between our method and the ground truth.

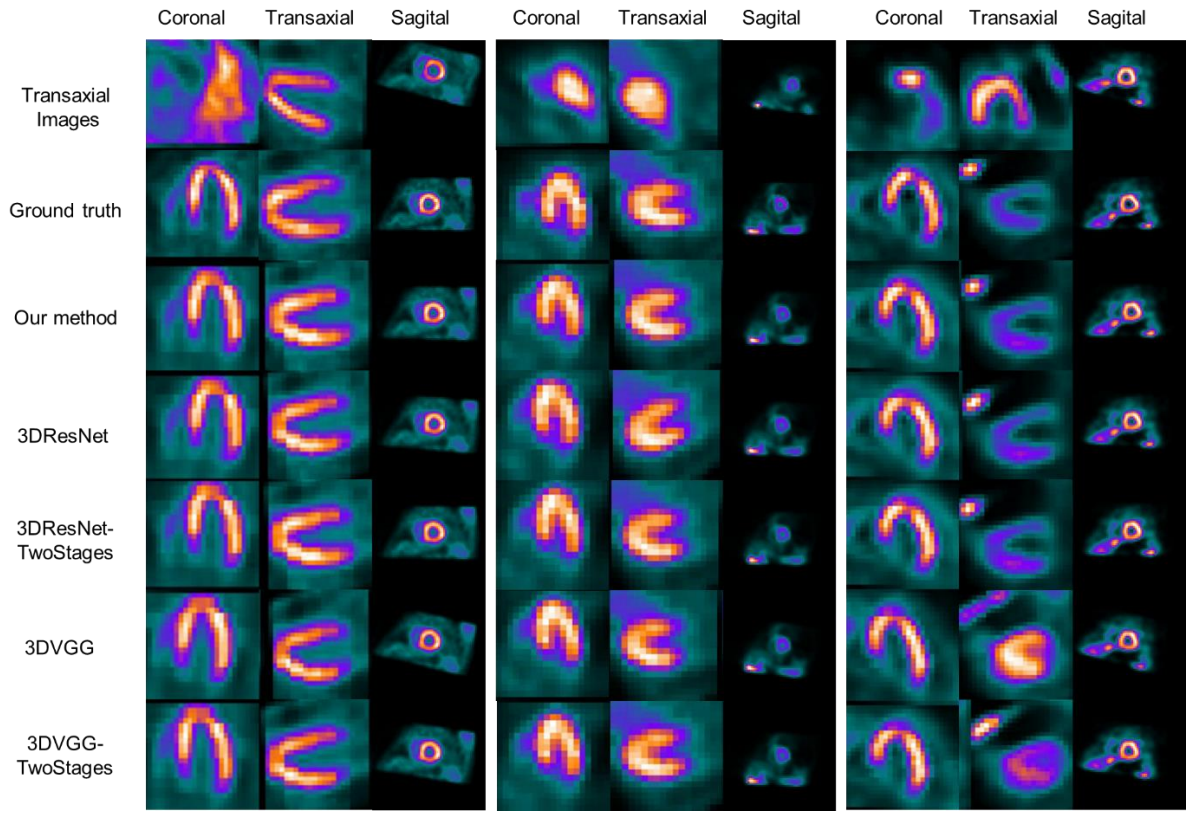

**FIGURE 4.** Transaxial images, reoriented images in the ground truth, and reoriented by different methods for three patients (left: normal, middle: small heart, right: abnormal uptake).

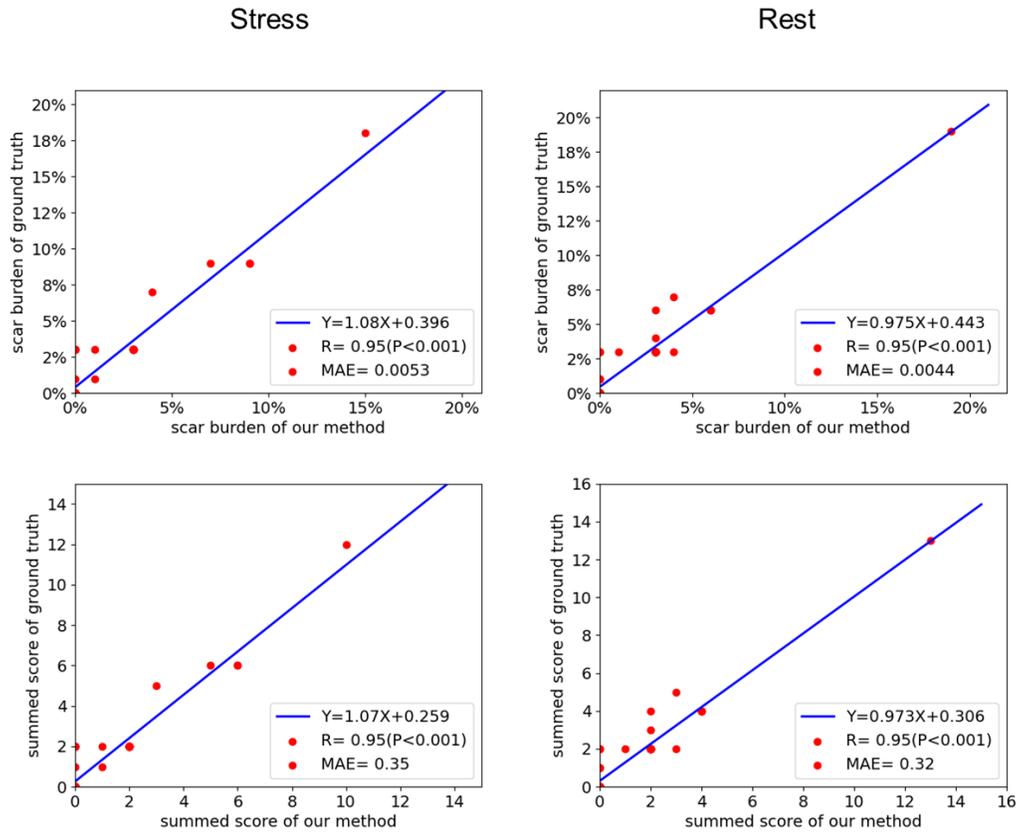

**FIGURE 5.** Linear regression analysis of scar burden and summed scores between our method and the ground truth for stress/ rest MPIs.

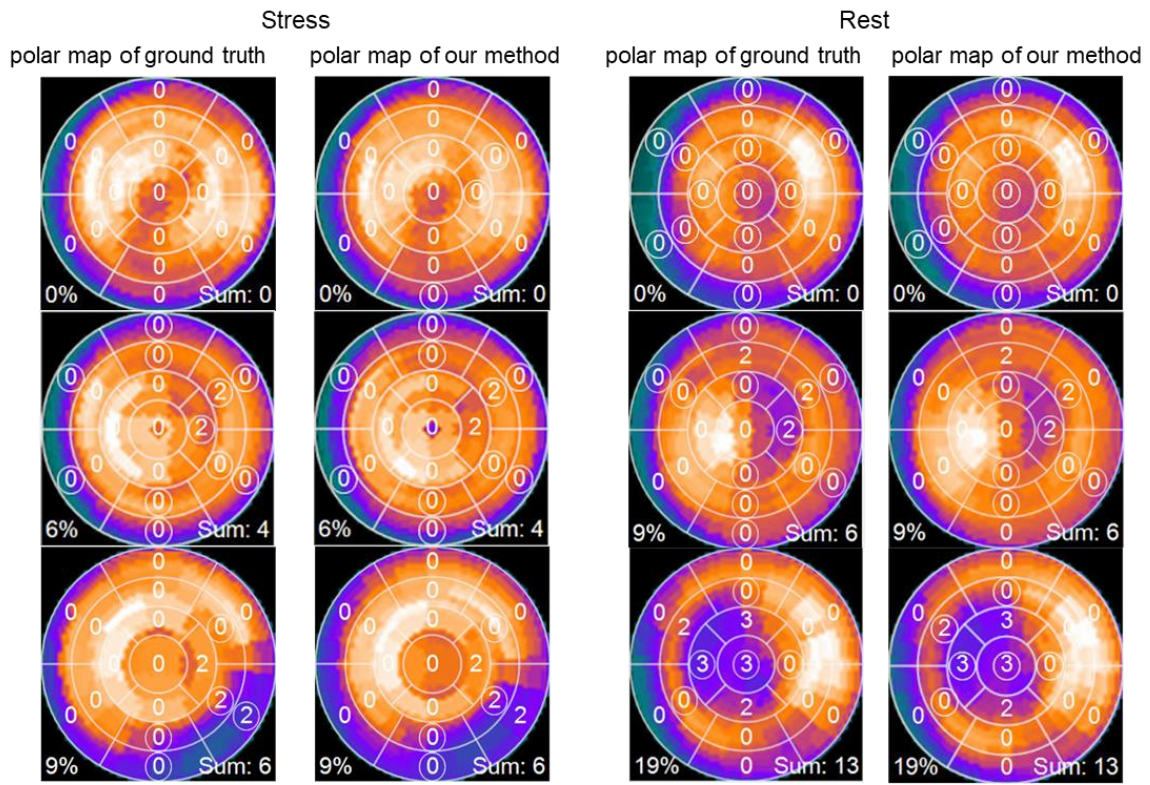

**FIGURE 6.** Polar maps of short-axis images generated from the ground truth and our method. The percentage at the left bottom of each polar map is the scar burden and the number at the right bottom is the summed score of perfusion defect. The number in each region of 17 segments is the score of segment perfusion defect.

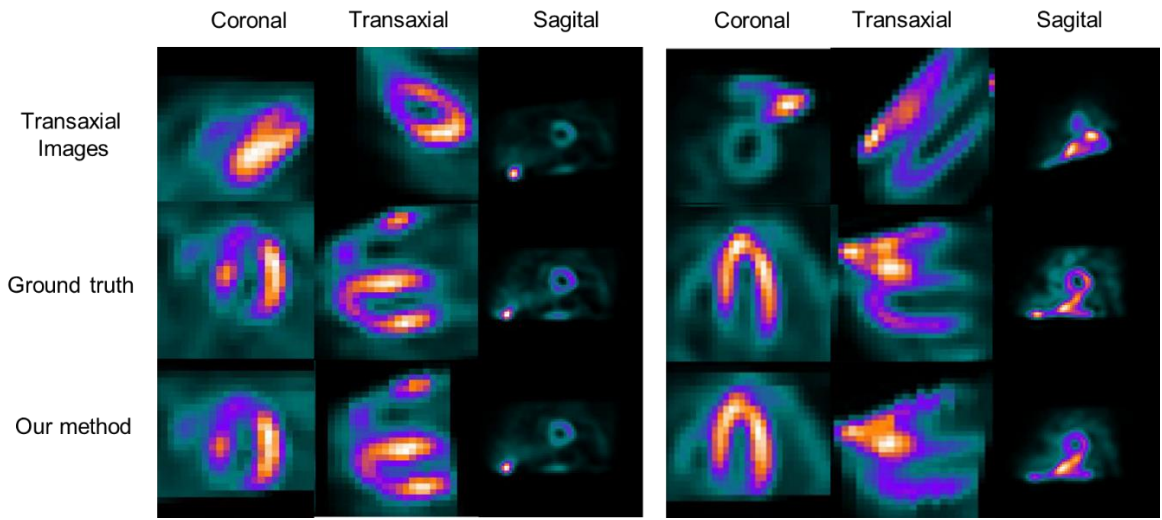

**FIGURE 7.** Results of two MPI cases reoriented by our deep-learning-based method. Left, an MPI with a rotation angle greater than 45 degrees; Right, an MPI with significant overlap between the left ventricle and organs.

Tables

**TABLE 1.** The accuracy (measured as the correlation coefficients of transformation parameters between prediction models and ground truth) and computational performance of different deep learning-based models for left-ventricular reorientation. 3DVGG and 3DResNet are 3D implementations of VGG and ResNet architectures. CNN+STN is the implementation of our method without the three-stage optimization. 3DVGG-ThreeStages and 3DResNet-ThreeStages are the modifications of 3DVGG and 3DResNet incorporating three-stage optimization strategy. X_ m, y_ m, z_ m are translation parameters; alpha, beta, theta are degrees of rotation parameters. Network complexity is the number of parameters in the network; training time measures the time consumption for each epoch (in minutes); prediction time is the time consumption for each MPI (in seconds). The bold texts indicate the best results.

|  | 3DVGG | 3DResNet | CNN+STN | 3DVGG-ThreeStages | 3DResNet-ThreeStages | Our method |
|---|---|---|---|---|---|---|
| x_m | 0.978 | 0.986 | 0.988 | 0.977 | **0.995** | 0.994 |
| y_m | 0.985 | 0.987 | 0.992 | 0.993 | 0.996 | **0.996** |
| z_m | 0.974 | 9.989 | 0.983 | 0.988 | **0.997** | 0.996 |
| alpha | 0.976 | 0.982 | 0.989 | 0.980 | 0.992 | **0.992** |
| beta | 0.988 | 0.991 | 0.993 | 0.990 | 0.996 | **0.997** |
| theta | 0.979 | 0.989 | 0.991 | 0.992 | 0.997 | **0.998** |
| Network complexity | 44,278,022 | 33,185,030 | **360,262** | 88,555,654 | 66,366,982 | 719,897 |
| Training time | 8.14 | 4.13 | **2.05** | 9.23 | 5.07 | 4.24 |
| Prediction time | 0.9426 | 0.1168 | **0.1109** | 1.7902 | 0.2253 | 0.2143 |